\documentclass[aps,prl,twocolumn]{revtex4-1}
\usepackage{amsmath}
\usepackage{amssymb}
\usepackage{graphicx}

\begin{document}

\title{What superconducts in sulfur hydrides under pressure, and why}
\author{N. Bernstein}
\author{C.~S. Hellberg}
\author{M.~D. Johannes}
\author{I.~I. Mazin}
\author{M.~J. Mehl}
\affiliation{Center for Computational Materials Science, Naval Research Laboratory,
Washington, DC}

\begin{abstract} 
The recent discovery of superconductivity at 190~K in highly
compressed H$_{2}$S is spectacular not only because it sets a record
high critical temperature, but because it does so in a material
that appears to be, and we argue here that it is, a conventional
strong-coupling BCS superconductor.
Intriguingly, superconductivity in the observed
pressure and temperature range was predicted theoretically in a
similar compound H$_{3}$S.  Several important questions about this
remarkable result, however, are left unanswered: (1) Does the
stoichiometry of the superconducting compound differ from the nominal
composition, and could it be the predicted H$_{3}$S compound? (2)
Is the physical origin of the anomalously high critical temperature
related only to the high H phonon frequencies, or does strong
electron-ion coupling play a role?  
We show that at experimentally
relevant pressures H$_2$S is unstable, decomposing into H$_3$S and S, and
that H$_3$S has a record high $T_c$ due to its covalent bonds driven
metallic.  The main reason for this extraordinarily high $T_c$ in H$_3$S
as compared with MgB$_2$, another compound with a similar superconductivity
mechanism, is the high vibrational frequency of the much lighter H
atoms.
\end{abstract}

\date{Printed on \today }
\pacs{Pacs Numbers: }
\maketitle

Recently reported superconductivity at 190~K in compressed H$_{2}$S~\cite{Eremets} has been arguably the biggest discovery in the 
field since the superconducting cuprates nearly thirty years earlier. Superconductivity in a related compound, H$_{3}$S, in a 
similar range of pressure with very nearly the same critical temperature 
was predicted theoretically~\cite{Cui} about a year 
earlier.  In that theoretical paper, direct \emph{ab initio} calculations yielded high phonon frequencies, giving the logarithmic 
average (the pre-factor in the equation for $T_{c}$) on the order of 1100-1300~K and a coupling constant $\lambda$ larger than 2, 
combining to give $T_c$ between 191-204~K. However, a microscopic understanding of why this particular material features such a 
strong coupling is still missing, as is an explanation of the discrepancy between experimental and theoretical stoichiometries.  As 
we show here, the answer lies in the stability of the H$_n$S series of compounds. At high pressure, the phase diagram favors 
decomposition of H$_2$S into H$_3$S and pure S.  The mechanism of superconductivity can be traced to the strongly covalent 
metallic nature of H$_3$S along with high phonon frequences, similar to another conventional (i.e.\ phonon-driven) superconductor 
with what now seems only a {\it relatively} high $T_c$, MgB$_2$~\cite{MgB2review}.

The discovery in 2001 of phonon-driven superconductivity at 39~K in MgB$_{2}$ not only set a record high $T_c$ for a conventional 
phonon-mediated mechanism, which just 30 years back was widely believed to be limited to $\lesssim 25$~K, but also introduced a completely new 
concept in the theory of superconducting materials, dubbed \textquotedblleft doped covalent bonds\textquotedblright\ by W.~E. Pickett and his 
collaborators~\cite{WEP}. The essence of this concept is that bonding and antibonding states in a covalent system are very sensitive to hopping 
integrals and thus to ionic positions, which makes them strongly coupled to the corresponding phonons. However, in essentially all covalent 
systems, the corresponding states are removed from the Fermi surface and thus irrelevant to superconductivity. Moreover, even when it is 
possible to dope a covalent insulator, such as diamond, it costs a tremendous amount of energy and therefore results in only very small doping 
levels. MgB$_{2}$ is different in two ways: First, it sports metallic $\pi $ bands in addition to the strongly covalent $\sigma $ bands; second, 
the $\sigma $ bands are 2D, and therefore even a small carrier concentration in these bands creates a sizeable density of states (DOS), that is, 
a substantial covalent metallicity. On the contrary, in diamond all the bands are 3D, In contrast, all the bands are 3D in diamond, and small doping 
causes a similarly small DOS, and thus a low critical temperature $T_{c}$.

While the general concept that hard phonons in H-rich materials might make them good superconductors is not new, so far 
high-$T_{c}$ superconductivity has been elusive, primarily because such hard phonons normally do not produce large coupling 
constants. As V.~L. Ginzburg wrote in 1977, \textquotedblleft \ldots in many already-known materials, the Debye temperature is very 
large, $\sim 10^{3}$~K, and low $T_{c}$ is related to small coupling constants \ldots In view of this, attention is attracted to 
various hydrogen-rich materials under high pressure.\textquotedblright \cite{VL} The strong electron-phonon coupling inherent to a 
covalent metal such as MgB$_{2}$ suggests a way to circumvent this problem, by combining metallized covalent bonds with lighter 
elements and higher phonon frequencies.  We argue that H$_3$S applies precisely this recipe to achieve its record critical 
temperature and that the physics of superconductivity, and, to some extent, even numerics, in H$_{3}$S is extremely similar to 
MgB$_{2}$ with the only qualitative difference being the factor of 11 between the masses of H and B. In plain English, H$_{3}$S is 
like MgB$_{2}$, but lighter.

To gauge the level of electron-phonon coupling, we note that the same electron-phonon interactions that contribute to phonon-mediated 
superconductivity also manifest themselves as the screening of bond-stretching force constants and softening of vibrations; this softening can 
therefore be taken as a proxy for the strength of the coupling. Recalling the case of MgB$_{2}$, we observe that it has covalent-bond stretching 
phonons at about 500~cm$^{-1}$ near the zone center, while in its sister compound AlB$_{2}$, which has its $\sigma $ band %away far from the 
Fermi level and uncoupled from conducting electrons, those phonons are as hard as 900~cm$^{-1}$~\cite{AlB2}. A softening of similar magnitude 
occurs in solid H$_{3}$S as compared with vibrations in S-H containing thiol molecules. The latter have frequencies of about 2500~cm$^{-1}$, 
while the calculated phonon frequencies in H$_{3}$S are roughly $\sim 1600$~cm$^{-1}$~\cite{Cui}. This is a very large softening indicating a 
very large coupling, even though it may not initially perceived as such because the bare frequency involving the light H atoms is so high.

In any theoretical analysis of a new material, it is important to establish the stoichiometry and the crystal structure of the 
compound of interest. For instance, a recently suggested theory~\cite{Hirsch} is based upon the H$_{2}$S composition, which, as we 
show below, is nearly certainly not the composition that supports superconductivity in the experiment. While the experiments showing 
superconductivity at 190~K started by compressing H$_2$S, they also showed the formation of pure S, suggesting that the material 
that actually superconducts is likely to be H enriched~\cite{Eremets}. The composition that Duan \emph{et al.}~\cite{Cui} studied 
theoretically, H$_{3}$S, is consistent with this observation, but they did not consider the full range of compositions in the 
phase diagram.  Previous theoretical publications have verified that the proposed high pressure phases for H$_{3}$S in 
Ref.~\onlinecite{Cui} and H$_{2}$S in Ref.~\onlinecite{Ma} are stable against decomposition into H and S. However, lack of 
decomposition into elemental species is not a particularly stringent test, and stability against separation into other phases, 
e.g.\ H$_{2}$S into S and H$_{3}$S, which is important for understanding the relevance of any calculations to the experiments in 
Ref.~\onlinecite{Eremets}, has not been investigated until now.

We begin by checking the stability of H$_n$S compounds with respect to decomposition into other phases by calculating their zero 
temperature enthalpy $H=E+PV$ as a function of composition using density functional theory calculations.  We used the VASP density functional 
theory software~\cite{kresse96} with the Perdew-Burke-Enzerhoff generalized gradient approximation~\cite{perdew96}, and a projector-augmented 
waves basis~\cite{blochl94,kresse99} with a 1000 eV~plane wave cutoff. The calculations used $16 \times 16 \times 16$ Monkhorst-Pack $k$-points 
for cells containing a single formula unit; correspondingly reduced $k$-point grids were used for the larger cells.  Geometries were relaxed 
under an applied pressure using the conjugate-gradient algorithm applied to both unit cell size and shape, as well as atomic positions, until 
the residual forces were less than 0.01~eV/{\AA}.  Since the stability of each composition depends on the enthalpy of the {\em lowest enthalpy} 
structure at that composition, we considered many structures at compositions ranging from pure S, to H$_n$S for $n=1-6$, to pure H\@.  The 
initial structures for our relaxation procedure came from previously published experimental and computational 
studies~\cite{luo93:S162GPa,degtyareva05:novelVI,pickard07,Cui,Ma}, manual modifications of these published structures, and from a simple 
version of the random structure search method~\cite{PickardRSS}. The final relaxed structures are listed in the Supplemental 
Material~\cite{supplemental}.

\begin{figure}
\centerline{\includegraphics[width=\columnwidth]{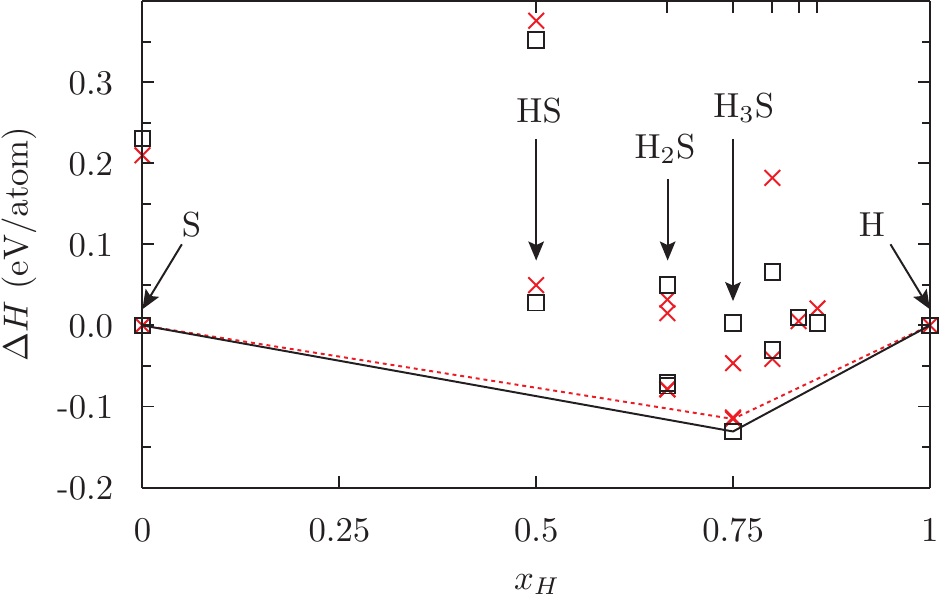}}
\caption{(color online) Formation enthalpy of H$_x$S$_{1-x}$ as a function of H concentration 
at $P=150$~GPa (red $\times$'s) and $200$~GPa (black squares).  Upper axis tick marks indicate compositions equivalent
to H$_n$S for integer $n$.  All compounds above the convex hull (lines) are unstable
with respect to decomposition into the two adjacent phases on the convex hull.
At both pressures the only stable compound is H$_3$S\@.
Pure S structures are taken from Refs.~\onlinecite{luo93:S162GPa} and 
\onlinecite{degtyareva05:novelVI}, and pure H from Ref.~\onlinecite{pickard07}. }
\label{fig:enthalpy}
\end{figure}

The zero temperature formation enthalpy of H$_x$S$_{1-x}$, $\Delta H(x)=H(x)-xH_{\rm H}-(1-x)H_{\rm S}$ as a function of $x$, is plotted in 
Fig.~\ref{fig:enthalpy}; we see that at $P=200$~GPa, H$_3$S in the previously proposed body-centered-cubic (bcc) $Im\bar{3}m$ structure is 
stable with respect to decomposition into any of the other calculated structures.  At 200~GPa the $R3m$ structure has cell and internal 
parameters that make it identical to the $Im\bar{3}m$, but below about 150~GPa it begins to continuously evolve to a distinct, lower symmetry 
structure~\cite{supplemental}.  All simulated H$_2$S structures, on the 
other hand, are unstable with respect to decomposition into S and H$_3$S by at least 120~meV/formula unit.  All structures with higher H 
concentration lie above the convex hull, and are therefore also unstable with respect to decomposition into H$_3$S and H.  In fact, all such 
structures include H atom pairs with distances similar to that of the H$_2$ molecule~\cite{supplemental}, showing the tendency to phase 
separation and release of pure H at these compositions.  Note that small errors in the enthalpies of the end points (pure S and pure H), if for 
example the true structures are a bit lower in enthalpy than the ones we considered, will not change these conclusions with respect to the 
stability of H$_3$S and instability of all other compositions. However, it is in principle possible that a sufficiently lower enthalpy H 
structure could make all intermediate compositions unstable, or that a H$_2$S structure with enthalpy outside of the convex hull, which has not 
been considered here, could exist. We think that the existence of such a lower enthalpy structure is very unlikely for pure H, which has been 
studied extensively~\cite{pickard07}. For H$_2$S, which is chemically reasonable and has not been previously studied extensively under such high 
pressure, we think that the failure of our thorough intuition-guided and random searches to find such a lower enthalpy structure makes its 
existence highly unlikely.  Finally, higher H concentration phases show a tendency for formation of H$_2$ molecules consistent with their 
predicted instability with respect to decomposition into pure H and H$_3$S\@.  Reducing the pressure to 150~GPa does not change the enthalpies 
of the low-lying structures significantly. We therefore conclude that under experimentally relevant pressures the starting material H$_2$S 
decomposes as 
\begin{equation}
  3H_{2}S \rightarrow 2H_{3}S+S,  \label{R1}
\end{equation} 
with no other products.

An important issue we have not addressed so far is the reliability of projector-augmented wave (PAW) calculations at such 
compressions since, generally speaking, the available PAWs were designed and tested for much larger interatomic separations. To 
this end, we compared the energy difference of the thermodynamic reaction (\ref{R1}) with the energy difference calculated with an 
all-electron method, full-potential local orbitals (FPLO)~\cite{FPLO}. We found a formation energy of 105~meV/formula unit, within 
10\% of the VASP value, confirming that the latter calculations are reliable.

Having established that the material exhibiting superconductivity at 190~K
at $P\sim 200$~GPa is most likely the $Im\bar{3}m$ bcc-like structure of H$_{3}$S 
identified in Ref.~\onlinecite{Cui}, we analyze its electronic
structure and superconductivity. We do not aim at recalculating the exact
numbers for the coupling constants and phonon frequencies, but rather at
gaining insight into why the calculations of Ref.~\onlinecite{Cui} produced such
a large coupling and $T_{c}.$

\begin{figure}
\centerline{\includegraphics[angle=0,origin=c,width=\columnwidth]{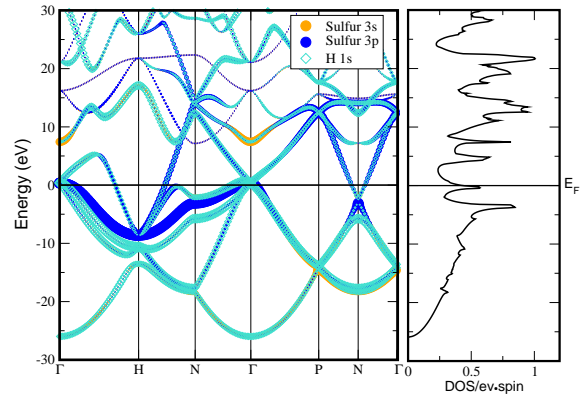}}
\caption{(color online) Band structure of the H$_3$S $Im\bar{3}m$ structure at 
$P=200$~GPa, calculated using FPLO. Weights of the three most important atomic
orbitals are shown with the symbols of the corresponding sizes.
The large contribution of H orbitals to the $S$-$s$ derived states at the bottom
of the band indicates strong covalency. }
\label{fig:band_struct}
\end{figure}

\begin{figure}
\centerline{\includegraphics[width=0.9\columnwidth]{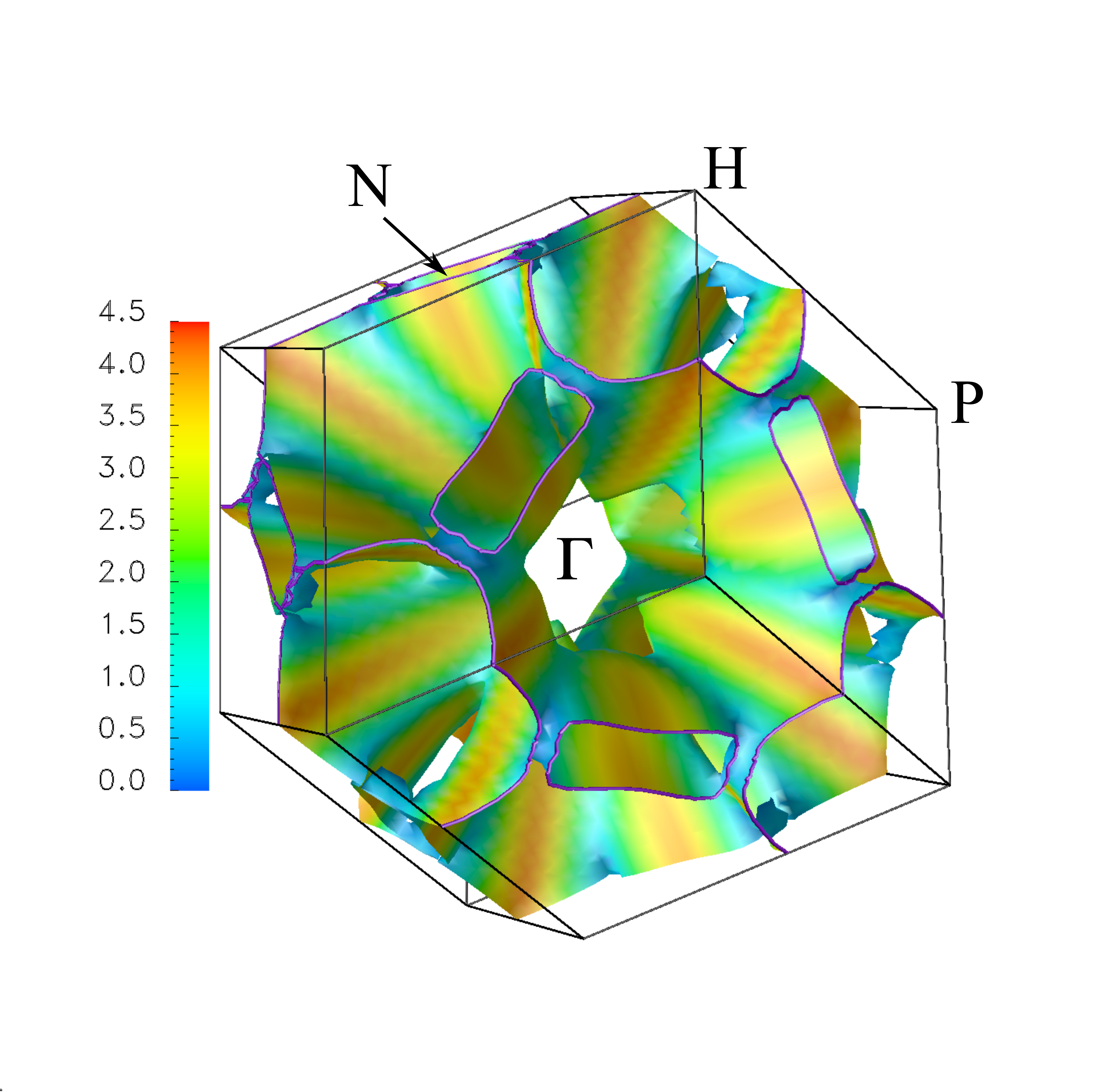}}
\caption{The main pocket of the Fermi surface of H$_3$S at 
$P=200$~GPa, colored according to the local Fermi velocity (the scale is in 
arbitrary units). Note heavy bands near the van Hove singularities. Two small hole 
pockets near $\Gamma$ and an electron pocket near $H$ contribute little to the 
total DOS, and are omitted from the picture.} 
\label{fig:fermi_surf}
\end{figure}

In Fig.~\ref{fig:band_struct} we show our calculated band structure, which
agrees with Duan \emph{et al.} 
The Fermi surface for the one band (out of the five that cross the Fermi level)
that contributes the overwhelming majority of the density of states, 
colored by the Fermi velocity, is shown in Fig.~\ref{fig:fermi_surf}.
The calculated average Fermi velocity, defined as 
$v_F=\sqrt{\Sigma _{\mathbf{k}}\delta (E_{\mathbf{k}}-E_{F})v_{\mathbf{k}}^{2}/\Sigma _{\mathbf{k}}\delta (E_{\mathbf{k}}-E_{F})}$ 
is $0.25 \times 10^{8}$~cm/sec. This allows us to
address another question: given the large $T_{c}$, would the coherence
length be long enough for the standard Eliashberg theory to be applicable? We
know that in high-$T_{c}$ cuprates this is nearly the case, which was argued
to have important theoretical implications in terms of Bose-condensation of local pairs 
rather than BCS long-range coherence\cite{BC}. For 
H$_3$S,
we can estimate the
zero temperature gap parameter, $\Delta (0)$, using Carbotte's formula~\cite{Carbotte}, $\Delta (0)=1.76k_{B}T_{c}[1+12.5(T_{c}/\omega _{\log })^{2}\log
(\omega _{\log }/2T_{c})]$. Using the numbers from Ref. \onlinecite{Cui}, we
get $\Delta (0)\approx 40$~meV. Using the standard expression for the clean
limit coherence length, $\xi =\hbar v_{F}/\pi \Delta (0),$ we find $\xi \sim
40$~{\AA}, much larger than the interatomic distance.

Analyzing the characters of the wave functions as in Fig.\ref{fig:band_struct}, we observe that the bands at
the Fermi level are formed nearly exclusively by seven orbitals: sulfur $3s$, 
sulfur $3p_{x,y,z}$, and the three $1s$ orbitals of the hydrogens, each displaced
along $x$, $y$, or $z$ from its nearest neighbor sulfur. The S $d-$states
are located more than 15 eV above the Fermi and can be safely neglected. In
the following we will denote the three hydrogen $1s$ orbitals by $i=x,y,z$,
and the three sulfur $p$ orbitals by $I=X,Y,Z$. The nearest neighbor S-H
Hamiltonian is 
\begin{eqnarray}
H_{si} &=&t_{a}C_{i}, \\
H_{Ii} &=&t_{b}S_{i},
\end{eqnarray}
where $t_{a,b}$ are the $ss\sigma $ and $ps\sigma $ S-H hoppings,
respectively, $C_{i}=2\cos (k_{i}a/2)$, and $S_{i}=2\sin (k_{i}a/2)$. The
H-H nearest neighbor hopping, despite this distance being the same as the
S-H one, is much smaller, 
due to the difference in radii of the H $1s$
and S $3s$ orbitals. The H-H Hamiltonian can be written down as 
\begin{equation}
H_{xy}=t_{c}\exp [i(k_{x}-k_{y})a/2]C_{z},
\end{equation}
etc. Using FPLO to construct these Wannier functions and their corresponding tight-binding Hamiltonian, we find $t_{a}=-4.2$~eV, 
$t_{b}= -5.2$~eV, and $t_{c}=-2.7$ {eV}. The onsite energies are: $E_{3s}=-8.6$~eV, $ E_{3p}=-1.3$ eV, and 
$E_{1s(H)}=-5.0$~eV ($E_{F}$ set to zero). 
We see that the hydrogen $s$ 
levels lie between the two sulfur levels, and create strong covalent bonds with both. The calculated bonding-antibonding splitting 
between sulfur $p$ and hydrogen $s$ states at the $P$ point ($ k=\{\pi /4a,\pi /4a,\pi /4a\})$ is $2\Delta \approx 25$~eV\@. 
However, the same splitting, by symmetry, is zero at the $\Gamma $ point, so, despite very strong covalency, this bond remains 
metallic, with the DOS $N(0)\sim 0.6 $~states/eV.

It is now instructive to compare these parameters with those in MgB$_{2}$,
keeping only bond-stretching boron phonons in MgB$_2$ and bond-stretching sulfur 
phonons in H$_3$S; in both cases these contributed about 70\% to the 
total coupling (it is worth noting that a 30\% contribution to the coupling 
constants does not imply a comparable contribution to $T_c$; in fact, since
$\omega_{\log}=\omega_S^{0.3}\omega_H^{0.7}$, excluding sulfur phonons 
would lead to very small changes in $T_c$). In order to do so, we use the well known qualitative relations
between the electron-phonon coupling constant $\lambda $, its electronic
part (also called Hopfield factor) $\eta $, and the average force constant $\Phi $. For the purpose of the ensuing discussion it is enough to know that
the Hopfield factor characterizes the electron-ion interaction and depends only
on electronic properties, such as the DOS and the ionic potential, but not
on phonon frequencies, while $\Phi $ is defined in such a way~\cite{VL2} that
it represents a combination of the derivatives of total energy with respect
to ionic coordinates (force constants) and thus carries information about
the phonon spectrum, but does not directly depend on the electronic energies
and wave functions. On a semiquantitative level~\cite{VL3},

\begin{eqnarray}
\lambda  &\approx &\eta /\Phi  \\
\Phi  &=&\Phi_{0}-2\eta .
\end{eqnarray}
Here $\Phi_{0}$ represents unscreened force constants that do not account
for electron-phonon coupling. Note that if there is one dominant phonon
mode, $\Phi_{0}/\Phi \approx \omega_{0}^{2}/\omega ^{2}$, where $\omega$
and $\omega_0$ are the screened and unscreened frequencies of this mode, 
respectively. Using the bond-stretching mode frequencies
for AlB$_{2}$ and MgB$_{2}$ to represent unscreened and screened bonds,
respectively, as mentioned above, we get 
\begin{eqnarray}
\Phi_{0}/\Phi  &\approx &(900/500)^{2}\sim 3 \\
\lambda  &\approx &\frac{1}{2}(\frac{\Phi_{0}}{\Phi }-1)\approx 1,
\end{eqnarray}
which qualitatively agrees with the accepted values for MgB$_{2}$~\cite{MgB2review},
$\lambda_{\sigma\sigma}=$0.78--1.02. This gives 
\begin{eqnarray}
\eta_{MgB_{2}} &\approx &1\times 11\times (500~\mathrm{cm}^{-1})^{2} \\
&\sim &2.75\times 10^{6}m_{H}\cdot \mathrm{cm}^{-2}.
\end{eqnarray}
If we assume the same Hopfield factor $\eta $ for H$_{3}$S, we get $\lambda
\approx $ $2.75\times 10^{6}/1300^{2}\approx 1.6$ (estimating the average
frequency of the bond-stretching modes in H$_{3}$S from Fig.~5 in Ref.~\onlinecite{Cui} 
as 40~THz), which is in the right ballpark compared to the
value $\lambda =2.19$ reported in Ref.~\onlinecite{Cui}. Note that the
logarithmic frequency in that paper was calculated to be 930~cm$^{-1}$,
suggesting that phonons softer than 40~THz contribute to total coupling.
Had we taken $\omega =1100$ cm$^{-1}=35$~THz instead of 40~THz, we would
have obtained the value calculated in Ref.~\onlinecite{Cui}. 

As a consistency check, let us now see whether this estimate implies a
plausible number for $\Phi_{0}$. Using $\omega =1300$~cm$^{-1}$ and $\lambda =1.6$, 
we find $\omega_{0}\approx 1300\sqrt{1+2\times 1.6}=2660$~cm$^{-1}$,
consistent with the vibron frequencies in S-H molecules, which have a large
gap and are unscreened by definition. The internal consistency of our
analysis and its consistency with experiment confirm our main conclusions
below.

In conclusion, at a pressure of $P=200$~GPa, H$_{2}$S gains nearly 40~meV per atom by decomposing into elemental 
sulfur and H$_{3}$S in the $Im\bar{3}m$ structure. The physical mechanism underlying the 
high-temperature superconductivity of H$_{3}$S at that pressure is very similar to that in MgB$_{2}$: 
metallization of covalent bonds. The main difference from MgB$_{2}$ is that the hydrogen mass is 11 
times smaller than the mass of boron, resulting in a 3.5 times larger prefactor.

We acknowledge useful discussions with D.~A. Papaconstantopoulos and M.
Calandra.

\end{document}

This supplement gives the crystallographic description of all of the
structures plotted in Figure 1 of the main text. As described in the
text, all calculations
were done with VASP, and the relative enthalpy
per atom, $\Delta H$, given for each structure is computed compared
to the sulfur $R\overline{3}m$ and hydrogen $C2/c$ endpoints.

\begin{table}[h]
  \caption{VASP minimum-enthalpy results for the high-pressure
    sulfur structures used in this paper. The parent structures are
    all obtained from the experimental
    literature.\cite{luo93:S162GPa,degtyareva05:novelVI}\label{tab:sulfur}}
  \begin{tabular}{cccc}
    \hline\hline
    Common name\cite{luo93:S162GPa} &
    \multicolumn{3}{c}{$\beta$Po} \\
    \hline
    \multicolumn{2}{c}{Space Group: $R\overline{3}m$ (\# 166)} & \multicolumn{2}{r}{Pearson Symbol: hR1} \\
    \hline\hline
    \multicolumn{1}{l}{150 GPa} & \multicolumn{3}{c}{$\Delta H$ = 0 } \\
    \hline
    $a, b, c$ & 2.15524\AA & 2.15524\AA & 2.15524\AA \\
    $\alpha~,~\beta~,~\gamma$ & 103.84707$^\circ$ & 103.84707$^\circ$ & 103.84707$^\circ$
    \\
    S (1a) & 0 & 0 & 0 \\
    \hline
    \multicolumn{1}{l}{200 GPa}  & \multicolumn{3}{c}{$\Delta H$ = 0
    }  \\
    \hline
    $a, b, c$ & 2.09715\AA & 2.09715\AA & 2.09715\AA \\
    $\alpha~,~\beta~,~\gamma$ & 104.17654$^\circ$ & 104.17654$^\circ$ & 104.17654$^\circ$
    \\
    S (1a) & 0 & 0 & 0 \\
    \hline\hline
    Common name\cite{degtyareva05:novelVI} &
    \multicolumn{3}{c}{S-III} \\
    \hline
    \multicolumn{2}{c}{Space Group: $I4_1/acd$ (\# 142)} & \multicolumn{2}{r}{Pearson Symbol: tI16} \\
    \hline\hline
    \multicolumn{1}{l}{150 GPa} & \multicolumn{3}{c}{$\Delta H$ =
      0.210 eV/atom} \\
    \hline
    $a, b, c$ & 7.46164\AA & 7.46164\AA & 2.61323\AA \\
    $\alpha~,~\beta~,~\gamma$ & 90$^\circ$ & 90$^\circ$ & 90$^\circ$
    \\
    S (16f) & 0.12449 & 0.37449 & 1/8 \\
    \hline
    \multicolumn{1}{l}{200 GPa} & \multicolumn{3}{c}{$\Delta H$ =
      0.230 eV/atom} \\
    \hline
    $a, b, c$ & 7.36018\AA & 7.36018\AA & 2.43068\AA \\
    $\alpha~,~\beta~,~\gamma$ & 90$^\circ$ & 90$^\circ$ & 90$^\circ$
    \\
    S (16f) & 0.12331 & 0.37331 & 1/8 \\
    \hline
  \end{tabular}
\end{table}

\begin{table}
  \caption{VASP minimum-enthalpy results for HS. 
%  The construction of the $C2/m$ phase is described in the main text. 
    \label{tab:HS}}
  \begin{tabular}{cccc}
    \hline\hline
    \multicolumn{2}{c}{Space Group: $C2/m$ (\# 12)} & \multicolumn{2}{r}{Pearson Symbol: mC16} \\
    \hline\hline
    \multicolumn{1}{l}{150 GPa}  & \multicolumn{3}{c}{$\Delta H$ =
      0.050 eV/atom}  \\
    \hline
    $a, b, c$ & 7.93061\AA & 2.79018\AA & 4.01968\AA \\
    $\alpha~,~\beta~,~\gamma$ & 90$^\circ$ & 105.21708$^\circ$ & 90$^\circ$
    \\
    H (4i) & 0.03276 & 0 & 0.73689 \\
    H (4i) & 0.18410 & 0 & 0.60249 \\
    S (4i) & 0.24626 & 0 & 0.20583 \\
    S (4i) & 0.46725 & 0 & 0.74690 \\
    \hline
    \multicolumn{1}{l}{200 GPa}  & \multicolumn{3}{c}{$\Delta H$ =
      0.028 eV/atom}  \\
    \hline
    $a, b, c$ & 7.65069\AA & 2.72463\AA & 3.91863\AA \\
    $\alpha~,~\beta~,~\gamma$ & 90$^\circ$ & 105.81312$^\circ$ & 90$^\circ$
    \\
    H (4i) & -0.03562 & 0 & 0.26699 \\
    H (4i) &  0.17759 & 0 & 0.59426 \\
    S (4i) &  0.24583 & 0 & 0.20267 \\
    S (4i) &  0.46321 & 0 & 0.74417 \\
    \hline\hline
    Common name &
    \multicolumn{3}{c}{NaCl} \\
    \hline
    \multicolumn{2}{c}{Space Group: $Fm\overline{3}m$ (\# 225)} & \multicolumn{2}{r}{Pearson Symbol: cF8} \\
    \hline\hline
    \multicolumn{1}{l}{150 GPa}  & \multicolumn{3}{c}{$\Delta H$ =
      0.376 eV/atom}  \\
    \hline
    $a, b, c$ & 3.49684\AA & 3.49684\AA & 3.49684\AA \\
    $\alpha~,~\beta~,~\gamma$ & 90$^\circ$ & 90$^\circ$ & 90$^\circ$
    \\
    S (4a) & 0 & 0 & 0 \\
    H (4b) & 1/2 & 1/2 & 1/2 \\
    \hline
    \multicolumn{1}{l}{200 GPa}  & \multicolumn{3}{c}{$\Delta H$ =
      0.352 eV/atom}  \\
    \hline
    $a, b, c$ & 3.40065\AA & 3.40065\AA & 3.40065\AA \\
    $\alpha~,~\beta~,~\gamma$ & 90$^\circ$ & 90$^\circ$ & 90$^\circ$
    \\
    S (4a) & 0 & 0 & 0 \\
    H (4b) & 1/2 & 1/2 & 1/2 \\
    \hline
  \end{tabular}
\end{table}

\begin{table}
  \caption{VASP minimum energy structures for H$_2$S. Except for the
    $C2/m$ structure, 
    %which is described in the main text, 
    all structures are from Ref.~\onlinecite{li14:H2S}. In the main text
  the structures are referred to by the International symbol for the
  space group name.\label{tab:H2S}}
  \begin{tabular}{cccc}
    \hline\hline
    \multicolumn{2}{c}{Space Group: $P\overline{1}$ (\# 2)} & \multicolumn{2}{r}{Pearson Symbol: aP6} \\
    \hline\hline
    \multicolumn{1}{l}{150 GPa}  & \multicolumn{3}{c}{$\Delta H$ =
      -0.079 eV/atom}  \\
    \hline
    $a, b, c$ & 4.07358\AA & 2.64522\AA & 2.64521\AA \\
    $\alpha~,~\beta~,~\gamma$ & 69.60083$^\circ$ & 103.32304$^\circ$ & 103.32279$^\circ$ \\
    H (1a) & 0 & 0 & 0 \\
    H (1g) & 0 & 1/2 & 1/2 \\
    H (2i) & 0.32665 & 0.33971 & 0.33968 \\
    S (2i) & 0.73684 & 0.18410 & 0.18410 \\
    \hline
    \multicolumn{1}{l}{200 GPa}  & \multicolumn{3}{c}{$\Delta H$ =
      -0.071 eV/atom}  \\
    \hline
    $a, b, c$ & 3.96516\AA & 2.56659\AA & 2.56659\AA \\
    $\alpha~,~\beta~,~\gamma$ & 70.18465$^\circ$ & 103.23036$^\circ$ & 103.23039$^\circ$ \\
    H (1a) & 0 & 0 & 0 \\
    H (1g) & 0 & 1/2 & 1/2 \\
    H (2i) & 0.32476 & 0.34280 & 0.34279 \\
    S (2i) & 0.73396 & 0.18459 & 0.18459 \\
    \hline\hline
    \multicolumn{2}{c}{Space Group: $C2/m$ (\# 12)} & \multicolumn{2}{r}{Pearson Symbol: mC12} \\
    \hline\hline
    \multicolumn{1}{l}{150 GPa}  & \multicolumn{3}{c}{$\Delta H$ =
      -0.079 eV/atom}  \\
    \hline
    $a, b, c$ & 4.34497\AA & 3.01695\AA & 4.07658\AA \\
    $\alpha~,~\beta~,~\gamma$ & 90$^\circ$ & 106.28258$^\circ$ & 90$^\circ$ \\
    H (2a) & 0 & 0 & 0 \\
    H (2b) & 0 & 1/2 & 0 \\
    H (4i) & 0.83989 & 1/2 & 0.32651 \\
    S (4i) & 0.68415 & 1/2 & 0.73691 \\
    \hline
    \multicolumn{1}{l}{200 GPa}  & \multicolumn{3}{c}{$\Delta H$ =
      -0.071 eV/atom}  \\
    \hline
    $a, b, c$ & 4.20060\AA & 2.95089\AA & 3.96490\AA \\
    $\alpha~,~\beta~,~\gamma$ & 90$^\circ$ & 106.25999$^\circ$ & 90$^\circ$
    \\
    H (2a) & 0 & 0 & 0 \\
    H (2b) & 0 & 1/2 & 0 \\
    H (4i) & 0.84277 & 1/2 & 0.32473 \\
    S (4i) & 0.68456 & 1/2 & 0.73390 \\
    \hline\hline
    \multicolumn{2}{c}{Space Group: $Cmca$ (\# 64)} & \multicolumn{2}{r}{Pearson Symbol: oC24} \\
    \hline\hline
    \multicolumn{1}{l}{150 GPa}  & \multicolumn{3}{c}{$\Delta H$ =
      -0.079 eV/atom}  \\
    \hline
    $a, b, c$ & 3.03134\AA & 4.39998\AA & 7.68189\AA \\
    $\alpha~,~\beta~,~\gamma$ & 90$^\circ$ & 90$^\circ$ & 90$^\circ$ \\
    H (8f) & 1/2 & 0.87315 & 0.77065 \\
    H (8f) & 1/2 & 0.10228 & 0.59070 \\
    S (8f) & 1/2 & 0.87240 & 0.11595 \\
    \hline
    \multicolumn{1}{l}{200 GPa}  & \multicolumn{3}{c}{$\Delta H$ =
      -0.074 eV/atom}  \\
    \hline
    $a, b, c$ & 2.97062\AA & 4.25243\AA & 7.44962\AA \\
    $\alpha~,~\beta~,~\gamma$ & 90$^\circ$ & 90$^\circ$ & 90$^\circ$ \\
    H (8f) & 1/2 & 0.36969 & 0.77078 \\
    H (8f) & 1/2 & 0.59714 & 0.59283 \\
    S (8f) & 1/2 & 0.37001 & 0.11422 \\
    \hline
  \end{tabular}
\end{table}

\begin{table}
  \caption{VASP minimum energy structures for H$_2$S. All structures
    are from Ref.~\onlinecite{li14:H2S}. In the main text the
    structures are referred to by the International symbol for the
    space group name. This is a continuation of
    Table~\ref{tab:H2S}. \label{tab:H2S_1}}
  \begin{tabular}{cccc}
    \hline\hline
    \multicolumn{2}{c}{Space Group: $Pc$ (\# 7)} & \multicolumn{2}{r}{Pearson Symbol: mP12} \\
    \hline\hline
    \multicolumn{1}{l}{150 GPa}  & \multicolumn{3}{c}{$\Delta H$ =
      0.015 eV/atom}  \\
    \hline
    $a, b, c$ & 4.32303\AA & 2.69863\AA & 4.58341\AA \\
    $\alpha~,~\beta~,~\gamma$ & 90$^\circ$ & 89.99963$^\circ$ & 90$^\circ$ \\
    H (2a) & 0.50267 & 0.05623 & 0.57855 \\
    H (2a) & 0.00266 & 0.46522 & 0.85143 \\
    H (2a) & 0.75075 & 0.15368 & 0.19896 \\
    H (2a) & 0.25458 & 0.84632 & 0.69896 \\
    S (2a) & 0.00267 & 0.16465 & 0.60343 \\
    S (2a) & 0.50267 & 0.41080 & 0.37567 \\
    \hline
    \multicolumn{1}{l}{200 GPa}  & \multicolumn{3}{c}{$\Delta H$ =
      0.050 eV/atom}  \\
    \hline
    $a, b, c$ & 4.16773\AA & 2.58857\AA & 4.39722\AA \\
    $\alpha~,~\beta~,~\gamma$ & 90$^\circ$ & 90.00247$^\circ$ & 90$^\circ$
    \\
    H (2a) & 0.50410 & 0.06567 & 0.70509 \\
    H (2a) & 0.00406 & 0.56569 & 0.89726 \\
    H (2a) & 0.75195 & 0.25217 & 0.04890 \\
    H (2a) & 0.25196 & 0.75215 & 0.55341 \\
    S (2a) & 0.00197 & 0.11884 & 0.66059 \\
    S (2a) & 0.50196 & 0.38114 & 0.44175 \\
    \hline\hline
    \multicolumn{2}{c}{Space Group: $Pmc2_1$ (\# 26)} & \multicolumn{2}{r}{Pearson Symbol: oP12} \\
    \hline\hline
    \multicolumn{1}{l}{150 GPa}  & \multicolumn{3}{c}{$\Delta H$ =
      0.032 eV/atom}  \\
    \hline
    $a, b, c$ & 4.30271\AA & 2.65496\AA & 4.54452\AA \\
    $\alpha~,~\beta~,~\gamma$ & 90$^\circ$ & 90 $^\circ$ & 90$^\circ$ \\
    H (2a) & 0 & 0.57229 & 0.90177 \\
    H (2b) & 1/2 & 0.07229 & 0.70056 \\
    H (4c) & 3/4 & 1/4 & 0.05117 \\
    S (2a) & 0 & 0.11684 & 0.65589 \\
    S (2b) & 1/2 & 0.38316 & 0.44644 \\
    \hline
    \multicolumn{1}{l}{200 GPa}  & \multicolumn{3}{c}{$\Delta H$ =
      0.050 eV/atom}  \\
    \hline
    $a, b, c$ & 4.16463\AA & 2.58689\AA & 4.40303\AA \\
    $\alpha~,~\beta~,~\gamma$ & 90$^\circ$ & 90$^\circ$ & 90$^\circ$
    \\
    H (2a) & 0 & 0.56603 & 0.89757 \\
    H (2b) & 1/2 & 0.06603 & 0.70476 \\
    H (4c) & 3/4 & 1/4 & 0.05116 \\
    S (2a) & 0 & 0.11917 & 0.66079 \\
    S (2b) & 1/2 & 0.38083 & 0.44155 \\
    \hline
  \end{tabular}
\end{table}

\begin{table}
  \caption{VASP minimum energy structures for H$_3$S. All structures
    are from Ref.~\onlinecite{duan14:H3S}. In the main text the
    structures are referred to by the International symbol for the
    space group name. Note that within computational accuracy the
    $R3m$ structure is identical to the $Im\overline{3}m$ structure
    at 200 GPa.\label{tab:H3S}}
  \begin{tabular}{cccc}
    \hline\hline
    \multicolumn{2}{c}{Space Group: $R3m$ (\# 160)} & \multicolumn{2}{r}{Pearson Symbol: hR4} \\
    \hline\hline
    \multicolumn{1}{l}{150 GPa}  & \multicolumn{3}{c}{$\Delta H$ =
      -0.115 eV/atom}  \\
    \hline
    $a, b, c$ & 2.65860\AA & 2.65860\AA & 2.65860\AA \\
    $\alpha~,~\beta~,~\gamma$ & 109.47092$^\circ$ & 109.47092$^\circ$ & 109.47092$^\circ$ \\
    H (3b) & 0.51487 & 0.51487 & -0.01768 \\
    S (1a) & 0 & 0 & 0 \\
    \hline
    \multicolumn{1}{l}{200 GPa}  & \multicolumn{3}{c}{$\Delta H$ =
      -0.131 eV/atom}  \\
    \hline
    $a, b, c$ & 2.58507\AA & 2.58507\AA & 2.58507\AA \\
    $\alpha~,~\beta~,~\gamma$ & 109.46968$^\circ$ & 109.46968$^\circ$ & 109.46968$^\circ$ \\
    H (3b) & 0.50000 & 0.50000 & 0.00000 \\
    S (1a) & 0 & 0 & 0 \\
    \hline\hline
    \multicolumn{2}{c}{Space Group: $Im\overline{3}m$ (\# 229)} & \multicolumn{2}{r}{Pearson Symbol: cI8} \\
    \hline\hline
    \multicolumn{1}{l}{150 GPa}  & \multicolumn{3}{c}{$\Delta H$ =
      -0.114 eV/atom}  \\
    \hline
    $a, b, c$ & 3.06473\AA & 3.06473\AA & 3.06473\AA \\
    $\alpha~,~\beta~,~\gamma$ & 90$^\circ$ & 90$^\circ$ & 90$^\circ$ \\
    H (6b) & 1/2 & 0 & 0 \\
    S (2a) & 0 & 0 & 0 \\
    \hline
    \multicolumn{1}{l}{200 GPa}  & \multicolumn{3}{c}{$\Delta H$ =
      -0.131 eV/atom}  \\
    \hline
    $a, b, c$ & 2.98504\AA & 2.98504\AA & 2.98504\AA \\
    $\alpha~,~\beta~,~\gamma$ & 90$^\circ$ & 90$^\circ$ & 90$^\circ$ \\
    H (6b) & 1/2 & 0 & 0 \\
    S (2a) & 0 & 0 & 0 \\
    \hline\hline
    \multicolumn{2}{c}{Space Group: $Cccm$ (\# 66)} & \multicolumn{2}{r}{Pearson Symbol: oC64} \\
    \hline\hline
    \multicolumn{1}{l}{150 GPa}  & \multicolumn{3}{c}{$\Delta H$ =
      -0.047 eV/atom}  \\
    \hline
    $a, b, c$ & 7.49494\AA & 7.47639\AA & 4.41236\AA \\
    $\alpha~,~\beta~,~\gamma$ & 90$^\circ$ & 90$^\circ$ & 90$^\circ$ \\
    H (8g) & 0.54900 & 0 & 1/4 \\
    H (8i) & 0 & 0 & 0.33751 \\
    H (8l) & 0.39933 & 0.15249 & 0 \\
    H (8l) & 0.34151 & 0.89639 & 0 \\
    H (16m) & 0.32199 & 0.67952 & 0.24893 \\
    S (8l) & 0.23320 & 0.06884 & 0 \\
    S (8l) & 0.43337 & 0.73718 & 0 \\
    \hline
    \multicolumn{1}{l}{200 GPa}  & \multicolumn{3}{c}{$\Delta H$ =
      0.003 eV/atom}  \\
    \hline
    $a, b, c$ & 7.28186\AA & 7.25388\AA & 4.23518\AA \\
    $\alpha~,~\beta~,~\gamma$ & 90$^\circ$ & 90$^\circ$ & 90$^\circ$ \\
    H (8g) & 0.55023 & 0 & 1/4 \\
    H (8i) & 0 & 0 & 0.34187 \\
    H (8l) & 0.40048 & 0.14696 & 0 \\
    H (8l) & 0.34520 & 0.89767 & 0 \\
    H (16m) & 0.31526 & 0.68567 & 0.24789 \\
    S (8l) & 0.22921 & 0.06464 & 0 \\
    S (8l) & 0.43842 & 0.73469 & 0 \\
    \hline
  \end{tabular}
\end{table}

\begin{table}
  \caption{VASP minimum energy structures for H$_4$S.
  % The construction of these structures is described in the text.
    \label{tab:H4S}}
  \begin{tabular}{cccc}
    \hline\hline
    Composition & \multicolumn{3}{c}{H$_4$S} \\
    \hline
    \multicolumn{2}{c}{Space Group: $I4/mmm$ (\# 139)} & \multicolumn{2}{r}{Pearson Symbol: tI10} \\
    \hline\hline
    \multicolumn{1}{l}{150 GPa}  & \multicolumn{3}{c}{$\Delta H$ =
      -0.041 eV/atom}  \\
    \hline
    $a, b, c$ & 2.98991\AA & 2.98991\AA & 3.74154\AA \\
    $\alpha~,~\beta~,~\gamma$ & 90$^\circ$ & 90$^\circ$ & 90$^\circ$ \\
    H (4c) & 1/2 & 0 & 0 \\
    H (4e) & 0 & 0 & 0.37538 \\
    S (2a) & 0 & 0 & 0 \\
    \hline
    \multicolumn{1}{l}{200 GPa}  & \multicolumn{3}{c}{$\Delta H$ =
      -0.030 eV/atom}  \\
    \hline
    $a, b, c$ & 2.91069\AA & 2.91069\AA & 3.63988\AA \\
    $\alpha~,~\beta~,~\gamma$ & 90$^\circ$ & 90$^\circ$ & 90$^\circ$ \\
    H (4c) & 1/2 & 0 & 0 \\
    H (4e) & 0 & 0 & 0.37597 \\
    S (2a) & 0 & 0 & 0 \\
    \hline
    \multicolumn{2}{c}{Space Group: $I4_1/a$ (\# 88)} & \multicolumn{2}{r}{Pearson Symbol: tI20} \\
    \hline\hline
    \multicolumn{1}{l}{150 GPa}  & \multicolumn{3}{c}{$\Delta H$ =
      0.182 eV/atom}  \\
    \hline
    $a, b, c$ & 2.97455\AA & 2.97455\AA & 7.31373\AA \\
    $\alpha~,~\beta~,~\gamma$ & 90$^\circ$ & 90$^\circ$ & 90$^\circ$ \\
    H (16f) & -0.25242 & 0.34189 & 0.33569 \\
    S (4b) & 0 & 1/4 & 1/8 \\
    \hline
    \multicolumn{1}{l}{200 GPa}  & \multicolumn{3}{c}{$\Delta H$ =
      0.066 eV/atom}  \\
    \hline
    $a, b, c$ & 2.89934\AA & 2.89934\AA & 7.0140\AA \\
    $\alpha~,~\beta~,~\gamma$ & 90$^\circ$ & 90$^\circ$ & 90$^\circ$ \\
    H (16f) & -0.24903 & 0.33054 & 0.16949 \\
    S (4b) & 0 & 1/4 & 1/8 \\
    \hline
  \end{tabular}
\end{table}

\begin{table}
  \caption{VASP minimum energy structures for H$_5$S and H$_6$S. 
  %The construction of these structures is described in the text.
    \label{tab:H56S}}
  \begin{tabular}{cccc}
    \hline\hline
    Composition & \multicolumn{3}{c}{H$_5$S} \\
    \hline
    \multicolumn{2}{c}{Space Group: $Amm2$ (\# 38)} & \multicolumn{2}{r}{Pearson Symbol: oC12} \\
    \hline\hline
    \multicolumn{1}{l}{150 GPa}  & \multicolumn{3}{c}{$\Delta H$ =
      0.005 eV/atom}  \\
    \hline
    $a, b, c$ & 2.98515\AA & 3.02370\AA & 4.13090\AA \\
    $\alpha~,~\beta~,~\gamma$ & 90$^\circ$ & 90$^\circ$ & 90$^\circ$ \\
    H (2a) & 0 & 0 & 0.00000 \\
    H (2a) & 0 & 0 & 0.80184 \\
    H (2b) & 1/2 & 0 & -0.00225 \\
    H (4c) & 0.18826 & 0 & 0.30139 \\
    S (2b) & 1/2 & 0 & 0.54161 \\
    \hline
    \multicolumn{1}{l}{200 GPa}  & \multicolumn{3}{c}{$\Delta H$ =
      0.010 eV/atom}  \\
    \hline
    $a, b, c$ & 2.88040\AA & 2.93948\AA & 4.01107\AA \\
    $\alpha~,~\beta~,~\gamma$ & 90$^\circ$ & 90$^\circ$ & 90$^\circ$ \\
    H (2a) & 0 & 0 & 0.00000 \\
    H (2a) & 0 & 0 & 0.79329 \\
    H (2b) & 1/2 & 0 & -0.00462 \\
    H (4c) & 0.18557 & 0 & 0.29492 \\
    S (2b) & 1/2 & 0 & 0.54435 \\
    \hline\hline
    Composition & \multicolumn{3}{c}{H$_6$S} \\
    \hline
    \multicolumn{2}{c}{Space Group: $C2$ (\# 5)} & \multicolumn{2}{r}{Pearson Symbol: mC14} \\
    \hline\hline
    \multicolumn{1}{l}{150 GPa}  & \multicolumn{3}{c}{$\Delta H$ =
      0.021 eV/atom}  \\
    \hline
    $a, b, c$ & 5.21666\AA & 3.05690\AA & 3.06384\AA \\
    $\alpha~,~\beta~,~\gamma$ & 90$^\circ$ & 124.49772$^\circ$ & 90$^\circ$ \\
    H (4c) & 0.01132 & 0.45761 & 0.63434 \\
    H (4c) & 0.20977 & 0.64117 & 0.30930 \\
    H (4c) & 0.31019 & 0.41475 & 0.22338 \\
    S (2a) & 0 & 0.00000 & 0 \\
    \hline
    \multicolumn{1}{l}{200 GPa}  & \multicolumn{3}{c}{$\Delta H$ =
      0.003 eV/atom}  \\
    \hline
    $a, b, c$ & 5.10741\AA & 2.94868\AA & 4.10351\AA \\
    $\alpha~,~\beta~,~\gamma$ & 90$^\circ$ & 143.80685$^\circ$ & 90$^\circ$ \\
    H (4c) & 0.09982 & 0.40763 & 0.40957 \\
    H (4c) & 0.12247 & 0.63513 & -0.09538 \\
    H (4c) & 0.37410 & 0.45731 & 0.38917 \\
    S (2a) & 0 & 0.00000 & 0 \\
    \hline
  \end{tabular}
\end{table}

\begin{table}
  \caption{VASP minimum energy structures for phase III of solid
    hydrogen.\cite{pickard07:HIII}\label{tab:H2}}
  \begin{tabular}{cccc}
    \hline\hline
    \multicolumn{2}{c}{Space Group: $C2/c$ (\# 15)} & \multicolumn{2}{r}{Pearson Symbol: mC24} \\
    \hline\hline
    \multicolumn{1}{l}{150 GPa}  & \multicolumn{3}{c}{$\Delta H$ = 0
    }  \\
    \hline
    $a, b, c$ & 5.41808\AA & 3.09239\AA & 4.54771\AA \\
    $\alpha~,~\beta~,~\gamma$ & 90$^\circ$ & 142.37109$^\circ$ & 90$^\circ$ \\
    H (4e) & 0 & 0.10256 & 1/4 \\
    H (4e) & 0 & 0.34211 & 1/4 \\
    H (8f) & 0.23699 & 0.08287 & 0.24251 \\
    H (8f) & 0.34320 & 0.19642 & 0.22032 \\
    \hline
    \multicolumn{1}{l}{200 GPa}  & \multicolumn{3}{c}{$\Delta H$ = 0
    }  \\
    \hline
    $a, b, c$ & 5.21924\AA & 2.97247\AA & 4.37603\AA \\
    $\alpha~,~\beta~,~\gamma$ & 90$^\circ$ & 142.43319$^\circ$ & 90$^\circ$ \\
    H (4e) & 0 & 0.10210 & 1/4 \\
    H (4e) & 0 & 0.35224 & 1/4 \\
    H (8f) & 0.23227 & 0.07725 & 0.24382 \\
    H (8f) & 0.34267 & 0.19559 & 0.22023 \\
    \hline
  \end{tabular}
\end{table}

\bibliography{abbrev,highpressure}